\documentclass[journal]{IEEEtran}

\usepackage{psfrag}
\usepackage{subfigure}
\usepackage{url}
\usepackage{stfloats}
\usepackage{cite}
\usepackage{amsmath}
\usepackage{amssymb}
\usepackage{amsfonts}
\usepackage{graphicx}
\usepackage{fancybox}
\usepackage{color}
\usepackage{algorithm}
\usepackage{algorithmic}
\usepackage{multirow}
\usepackage{upgreek}
\usepackage{booktabs}
\usepackage{threeparttable}
\usepackage{latexsym}
\usepackage{bm,array}
\usepackage{graphicx}
\usepackage{float}
\usepackage{mathtools}
\usepackage{commath}

\usepackage{stfloats}
\usepackage{cuted}
\usepackage{amsmath,lipsum}
\usepackage{cuted}

\usepackage{fancyhdr}
\usepackage{kantlipsum}

\IEEEoverridecommandlockouts
\normalsize

\fancypagestyle{plain}{
  \fancyhf{}
  \fancyhead[C]{Conference on \LaTeX}     
  \fancyfoot[L]{This is a notice}

}
\usepackage{eso-pic}

\begin{document}
\AddToShipoutPictureBG*{%
  \AtPageUpperLeft{%
    \setlength\unitlength{1in}%
    \hspace*{\dimexpr 0.46\paperwidth\relax}
    \makebox(0,-0.75)[c]{\footnotesize This work has been submitted to the IEEE for possible publication. Copyright may be transferred without notice, after which this version may no longer be accessible}%
}}
\AddToShipoutPictureBG*{%
  \AtPageLowerLeft{%
    \setlength\unitlength{1in}%
    \hspace*{\dimexpr0.5\paperwidth\relax}
    \makebox(0,0.75)[c]{\Large Notice}%
}}


\title{Technology Trends for Massive MIMO towards 6G}

\author{
Yiming~Huo,~\IEEEmembership{Senior Member,~IEEE},
Xingqin Lin,~\IEEEmembership{Senior Member,~IEEE},
Boya Di,~\IEEEmembership{Member,~IEEE},
Hongliang Zhang,~\IEEEmembership{Member,~IEEE},
Francisco Javier Lorca Hernando,
Ahmet Serdar Tan,
Shahid Mumtaz,~\IEEEmembership{Senior Member,~IEEE},
Özlem Tuğfe Demir,~\IEEEmembership{Member,~IEEE},
and Kun Chen-Hu,~\IEEEmembership{Member,~IEEE}

\thanks{Yiming Huo is with the Department of Electrical and Computer Engineering, University of Victoria, Victoria, BC V8P 5C2, Canada (ymhuo@uvic.ca).}  

\thanks{Xingqin Lin is with NVIDIA, Santa Clara, CA 95050, USA (xingqinl@nvidia.com).}

\thanks{Boya Di, and Hongliang Zhang are with the School of Electronics, Peking University, Beijing 100871, China (e-mail: boya.di@pku.edu.cn; hongliang.zhang92@gmail.com).}

\thanks{Francisco Javier Lorca Hernando, and Ahmet Serdar Tan are with InterDigital Communications, Inc. London, England, United Kingdom (e-mail: javier.lorcahernando@interdigital.com; AhmetSerdar.Tan@interdigital.com).}

\thanks{Shahid Mumtaz is with the Instituto de Telecomunicações, Aveiro, Portugal (e-mail: smumtaz@av.it.pt).}

\thanks{Özlem Tuğfe Demir is with the Department of Computer Science, KTH Royal Institute of Technology, Stockholm, Sweden (e-mail: ozlemtd@kth.se).} 

\thanks{Kun Chen-Hu is  with the Department of Signal Theory and Communications of Universidad Carlos III de Madrid, Leganés, 28911, Spain (e-mail: kchen@tsc.uc3m.es).} 

}

\maketitle

\begin{abstract}
At the dawn of the next-generation wireless systems and networks, massive multiple-input multiple-output (MIMO) has been envisioned as one of the enabling technologies. With the continued success of being applied in the 5G and beyond, the massive MIMO technology has demonstrated its advantageousness, integrability, and extendibility. Moreover, several evolutionary features and revolutionizing trends for massive MIMO have gradually emerged in recent years, which are expected to reshape the future 6G wireless systems and networks. Specifically, the functions and performance of future massive MIMO systems will be enabled and enhanced via combining other innovative technologies, architectures, and strategies such as intelligent omni-surfaces (IOSs)/intelligent reflecting surfaces (IRSs), artificial intelligence (AI), THz communications, cell free architecture. Also, more diverse vertical applications based on massive MIMO will emerge and prosper, such as wireless localization and sensing, vehicular communications, non-terrestrial communications, remote sensing, inter-planetary communications.    

\end{abstract}

\begin{IEEEkeywords}
6G, Massive MIMO, Intelligent Omni-Surface (IOS), Intelligent Reflecting Surface (IRS), Cell Free, Artificial Intelligence, Vehicular Communications, THz Communications, Non-Terrestrial Communications, Remote Sensing, Inter-Planetary Communications. 

\end{IEEEkeywords}

\section{Introduction}


Massive multiple-input multiple-output (MIMO) has been one of the essential technologies in 5G wireless communications and recently experiencing unprecedented growth in development and deployment. With fast technological innovations and enormous commercial needs, mass MIMO is expected to evolve further and reshape future telecommunications and related areas more broadly and deeply.

Support for massive MIMO is intrinsic in 5G New Radio (NR) standards \cite{ref1}. As the first release of 5G NR, Release 15 includes the fundamental features to support massive MIMO in different deployment scenarios, including reciprocity-based operation for time division duplex (TDD) systems, high-resolution channel state information (CSI) feedback for multi-user MIMO (MU-MIMO), and advanced beam management for high-frequency band operation with analog beamforming, among others. After Release 15, 3GPP specified further enhancements of massive MIMO in Release 16. Representative massive MIMO enhancements in Release 16 are CSI feedback overhead reduction through spatial and frequency domain compression, beam management signaling overhead and latency reduction, and non-coherent joint transmission from multiple transmit and receive points (TRPs). 

3GPP continued massive MIMO evolution in Release 17. CSI feedback overhead was further reduced by exploiting angle-delay reciprocity. A unified transmission configuration indicator (TCI) framework was introduced to enhance multi-beam operation. Multi-TRP support was also improved with the introduction of inter-cell multi-TRP enhancements and multi-TRP-specific beam management features. Release 18 is the start of work on 5G Advanced, and its scope includes further massive MIMO evolution. Potential directions under investigation in 3GPP are uplink MIMO enhancements (e.g., the use of eight transmission antennas in the uplink and multi-panel uplink transmission), an extension of the unified TCI framework from single TRP to multi-TRP scenarios, a larger number of orthogonal demodulation reference signal (DMRS) ports for MU-MIMO, and CSI reporting enhancements for user equipment (UE) with medium and high velocities.

With fast standardization and promising commercialization, massive MIMO becomes the critical underlying technology for 5G and beyond, and is expected to combine other enabling technologies and expand to more new verticals. This article presents and analyzes several technology trends for massive MIMO evolving on the path to 6G. For example, one of the critical  observations is the recent intense attention paid to intelligent surfaces \cite{ref-IRS} which hold great potential to enable energy/cost-efficient massive MIMO. Furthermore, the intelligent reflecting surface (IRS) enabled massive MIMO can facilitate joint communications, localization and sensing functions that extensively enable new use cases and strengthen the wireless system performances in 6G. 

This article's first two sections are dedicated to IRS physical fundamentals for massive MIMO and IRS-enabled massive MIMO for localization and sensing, respectively. Immediately afterward, we provide a survey on ultra-massive MIMO at THz frequencies since adopting small wavelengths, and wide bandwidth brings unprecedented challenges in almost every aspect of the wireless system design. Then, we investigate the cell-free massive MIMO technology which improves spectral and energy efficiency. Next, the artificial intelligence (AI) for massive MIMO is surveyed and discussed, followed by a review and discussion of massive MIMO-OFDM for high-speed applications. As the last vertical of future massive MIMO towards 6G, non-terrestrial networks (NTNs) communications has been one direction in standardization since 2017 \cite{Lin-Space}. Thus, we present a detailed review, discussion, and analysis of current and futuristic non-terrestrial applications and architectures on top of massive MIMO before concluding this article.

\section{IOS/IRS Physical Fundamentals for Massive MIMO}  

Driven by the explosive growth in wireless data traffic, there is pressing need for innovative communication paradigms supporting high data rates in the future 6G. Massive MIMO has attracted heated attention exploiting the implicit randomness of the wireless environment. However, traditional massive MIMO relies on the large-scale phase arrays, which induce high hardware cost and power consumption due to the energy-consuming phase shifters, especially when the number of antennas grows. This limits their scalability to support massive MIMO in practice.  

Recently, intelligent metasurface, as a new type of ultra-thin two-dimensional metamaterial inlaid with sub-wavelength scatters, has provided a novel technology to enable massive MIMO in a cost-efficient way. Capable of reflecting and/or refracting the incident signals simultaneously, the surface can actively shape uncontrollable wireless environments into a desirable form via flexible phase shift reconfiguration \cite{ref-IOS}. Since such reconfiguration of each element is usually achieved via one or two PIN diodes controlled by the biased voltage, it only involves little hardware and power costs compared to the traditional phase arrays. Thus, the surface can be easily extended to a large scale, providing a practical method for realizing massive MIMO. 

\begin{figure}[t!]
	\centering
	\includegraphics[width=0.36\textwidth]{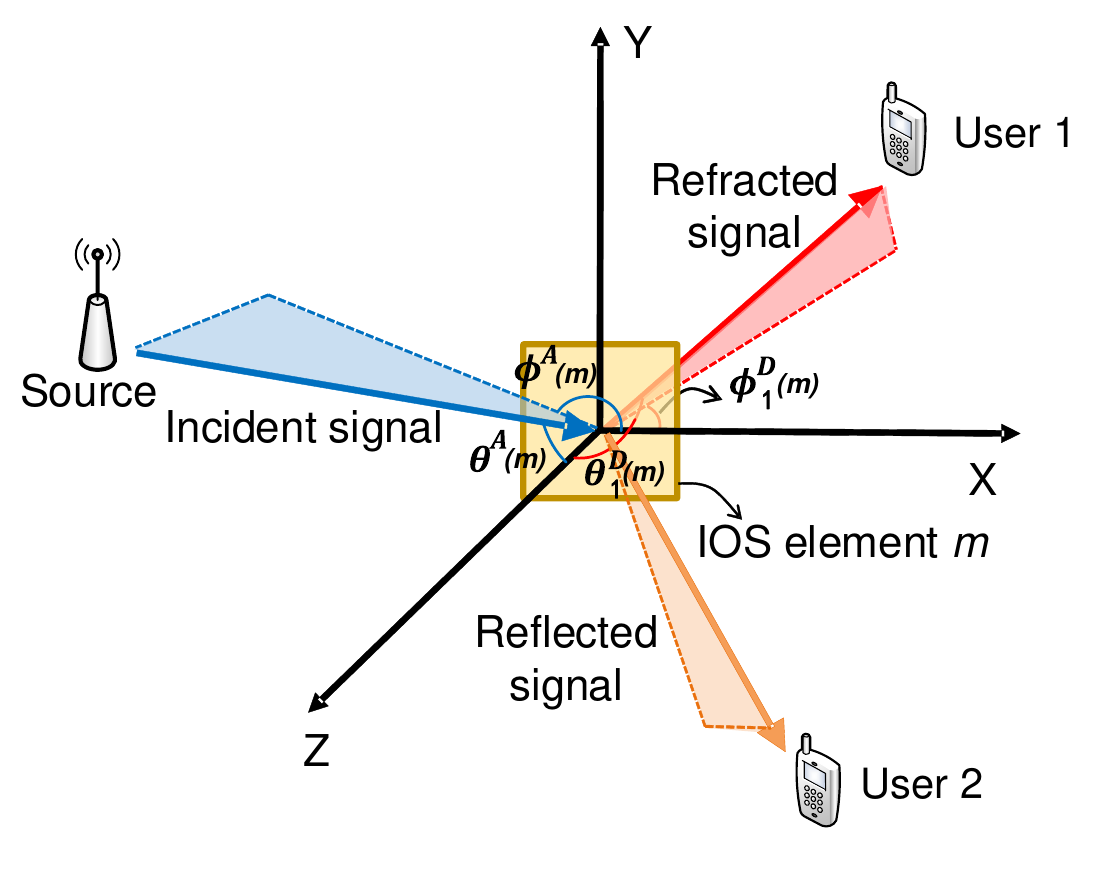}
	\caption{Transmission model of an intelligent surface element.}
	\label{fig:phy_fundamentals}
\end{figure}

A general transmission model of one surface element is shown in Fig.\ref{fig:phy_fundamentals}. After the incident signal arrives at the surface, part of it is reflected and the rest is refracted towards the other side. By defining the reflection-refraction ratio as $\epsilon$, we have the reflected and refracted signals in Fig.\ref{fig:phy_fundamentals}. Three different types of surfaces can then be classified below:
\begin{itemize}
	\item When $\epsilon = 0$, the surface only reflects the incident signal, leading to an intelligent reflecting surface. It can be attached to the wall, serving as a reflective relay for coverage.
	\item When $\epsilon \rightarrow \infty$, the surface only refracts the incident signal, serving as an reconfigurable refractive surface (RRS). It can replace the antenna array at the base station for transmission and reception. 
	\item When $0 < \epsilon < \infty$, the surface can reflect and refract the incident signal simultaneously, named as intelligent omni-directional surface (IOS). Compared to IRS, it can achieve full-dimensional wireless communications despite users's locations with respect to the surface. 
\end{itemize}

Both IOS and IRS have been considered as efficient methods to achieve massive MIMO due to their mature implementation. Especially, the recently developing IOS technique has also brought new challenges to the field:
\begin{itemize}
	\item The refracted and reflected signals of IOS are coupled with each other, determined simultaneously by the states of PIN diodes. Such a coupling effect makes it unknown whether IOS has the same impact on EM
	waves when the signal impinges on different sides of the IOS, i.e., whether the channel reciprocity still holds for the IOS-aided transmissions. 
	\item Besides, to fully exploit the refract-and-reflect characteristic of IOS, it is also necessary to explore the optimal position and orientation of the IOS given the BS and user distribution to extend the coverage on both sides of the IOS.
	\item In addition, a beamforming scheme should be reconsidered and tailored for the IOS-aided transmission since the reflected and refracted beams towards different users are dependent with each other \cite{ref-IOS-mag}. 
\end{itemize}

\section{Localization and Sensing Using IOS/IRS Enabled Massive MIMO}  

\begin{figure}[t!]
\centering
\includegraphics[width=0.4\textwidth]{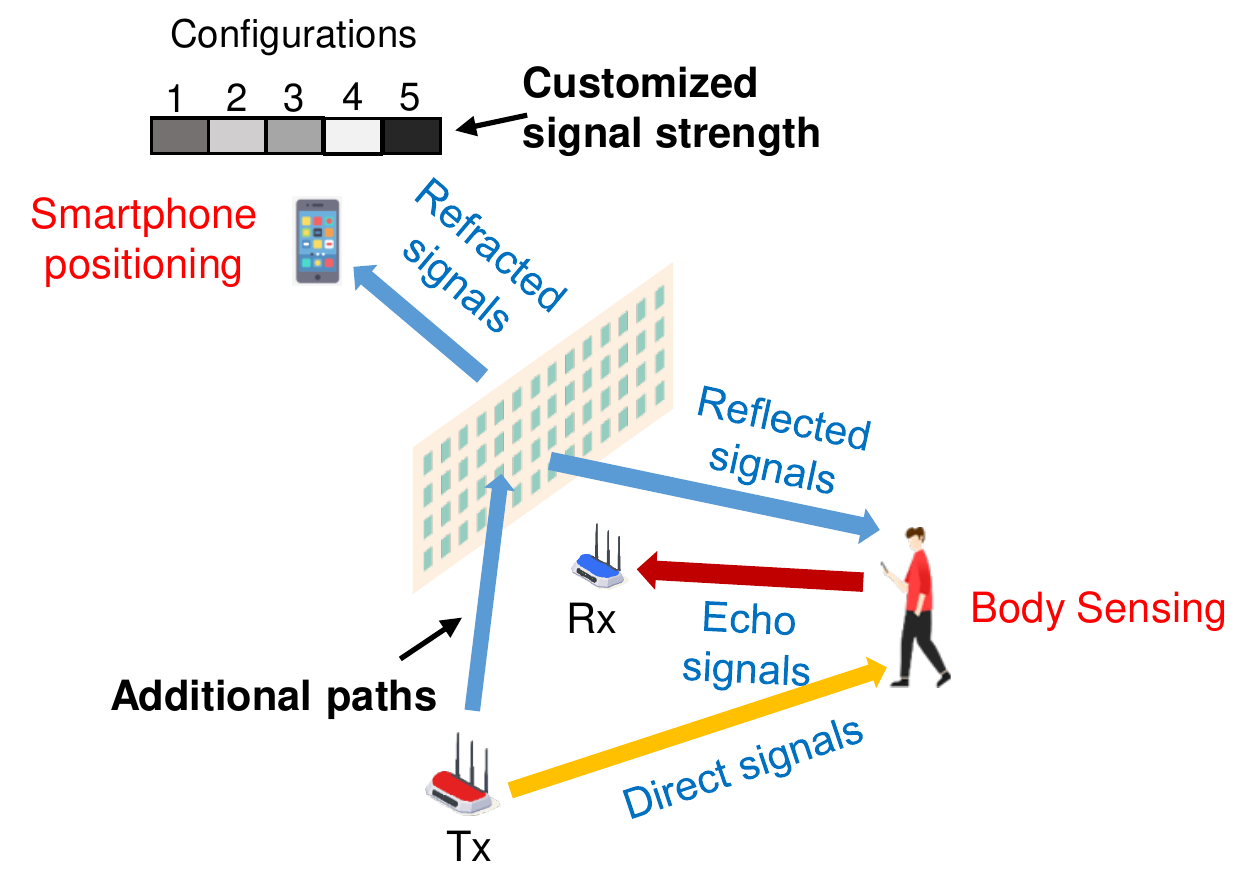}
\caption{Illustration for wireless localization and sensing using IOS enabled massive MIMO systems.}
\label{fig:sensing}
\end{figure}

In future 6G, wireless localization and sensing functions will be built-in for various applications, e.g., navigation, transportation, and healthcare. As a result, it is highly demanding to provide services with fine-resolution sensing and high localization accuracy. To realize this vision, massive MIMO can be a promising solution as the beam width can be reduced with a larger antenna array, leading to a high spatial resolution. However, the wireless environments in these systems are becoming complicated, for example, line-of-sight (LoS) links might be blocked by buildings or objects, degrading the accuracy of sensing and localization. Fortunately, the development of the IOS can provide favorable propagation conditions to improve sensing and localization accuracy \cite{ref-Sensing}. On the one hand, the IOS can provide additional paths toward targets, extending the coverage. On the other hand, with the capability of manipulating propagation conditions, the signals from different objects or targets can be customized so that they are easier to be distinguished, as illustrated in Fig. \ref{fig:sensing}.

Nevertheless, the integration of the IOS in a wireless sensing/localization system is not trivial, which generally brings the following challenges:
\begin{itemize}
    \item It will be a challenge to optimize the configurations relating to the IOS. Different from the designs of the IOS for the communication purposes, the optimizations here aim to maximize the sensing/localization performance, necessitating new designs. For example, the metric could be defined as the distance between two signal patterns (each corresponding to a configuration of the IOS) from different targets/positions so that the receiver could recognize two targets/positions with less efforts, leading to a higher accuracy. Moreover, as the number of IOS elements could be large, it will cause prohibitively high delay to enumerate all the configurations. Therefore, it will be important to select an appropriate number of configurations to achieve the trade-off between the latency and accuracy.
    
    \item The coupling of decision function with the optimization of the IOS makes it hard to find the optimal function. To be specific, the receiver needs a decision function to transform the received signals into the information of targets/positions. As the received signals can be adjusted by the IOS, the selection of decision function is also influenced by the configurations of the IOS. Therefore, a joint optimization will be necessary to improve the performance \cite{ref-Sensing}.

    \item In addition to the above signal processing challenges, practical implementation is another challenge. Where to deploy the IOS and how to determine its size should be carefully addressed, which should also take the topology of the environment into consideration. 
\end{itemize}

To sum up, massive MIMO technology will be expected to provide multi-functional services integrating communication, localization, and sensing. The IOS, which could customize the propagation environments, is believed to be an add-on enabler for future massive MIMO to facilitate such an integration.

\section{Ultra-Massive MIMO at THz Frequencies}  

According to ITU-R (International Telecommunication Union Radiocommunication Sector), THz frequencies are those in the range 0.1 THz – 10 THz. The lowest frequency region between 0.1 THz and 0.3 THz with the highest potential is usually called the sub-THz regime. THz and sub-THz signals serve as a bridge between radio and optical frequencies. Their wavelengths in the millimeter and sub-millimeter region make them excellent candidates to fulfill the 6G promise of extremely high-capacity communications, good situational awareness, and ultra-high resolution environmental sensing. Such small wavelengths come at the price of high uncertainty in the channel characteristics, leading to unreliable, intermittent radio links that suffer from one or several of the following impairments: 

\begin{itemize}
    \item High path losses, molecular absorption, and blockage. The high free-space path loss motivated by the small antenna aperture areas at these frequencies, together with the molecular absorption, blockage, diffuse scattering, and extra attenuation caused by rain, snow, or fog, lead to highly intermittent links. Link reliability must therefore be improved with the use of ultra-narrow beamforming.
    
    \item Low energy efficiency. RF output power degrades 20 dB per decade for a given Power Amplifier (PA) technology. This compromises the link budget and reinforces the need of large-scale transceivers with high numbers of antennas.
    
    \item Large-scale transceivers. The high beamforming gain needed to improve link reliability demands large-scale transceivers with a high number of antennas (usually, more than 1024). The sharpened, ultra-narrow beams that they produce pose significant challenges to mobility and beam tracking.
    
    \item Phase noise. At sub-THz/THz frequencies, CP-OFDM performance can be severely degraded by the inter-carrier interference (ICI) resulting from phase noise. Increasing the subcarrier spacing can mitigate its impact, but the correspondingly shorter symbol duration introduces a penalty in coverage and impairs the ability to mitigate large delay spreads.
    
    \item Channel sparsity. Ultra-narrow beams, together with ray-like wave propagation, lead to channels that exhibit small numbers of spatial degrees of freedom and ranks limited to one LoS component and a few multipath components, which challenges MIMO operation.
    
    \item Spherical wave and near-field effects. Large-scale transceivers exhibit significant spherical wave and near-field effects from the electrically large antenna structures that they equip, which introduces complexity to MIMO precoding strategies. 
    
    \item Beam squint. The narrowband response of phase shifters in planar arrays introduces a frequency-dependent beam misalignment called beam squint. Losses from beam misalignments can be alleviated by using beam broadening techniques, at the cost of reduced coverage; and avoided with true time delay (TTD) units, at the cost of complexity.
    
\end{itemize}

There is abundant research on transceiver architectures and network solutions aimed to ameliorate some of the above issues, especially those motivated by the propagation challenges at sub-THz/THz. Among the network solutions, the aforementioned IRS/RIS equipped with very large numbers of small antenna elements are receiving considerable attention, because of their ability to tailor the characteristics of the reflected and refracted beams \cite{ref-IRS}, \cite{ref-MIMO-Evolution}. IRS/RIS at sub-THz/THz exploit ray deflections to overcome blocking and path loss; can take benefit of the near-field effects by focusing beams to improve beamforming and 3D imaging; and can enhance the multipath richness of the channel to reinforce the spatial multiplexing capabilities at sub-THz/THz frequencies.

\section{Cell-Free Massive MIMO}  

\begin{figure}[t!]
\centering
\includegraphics[trim={0mm 1mm 2mm 4mm},clip,width=0.41\textwidth]{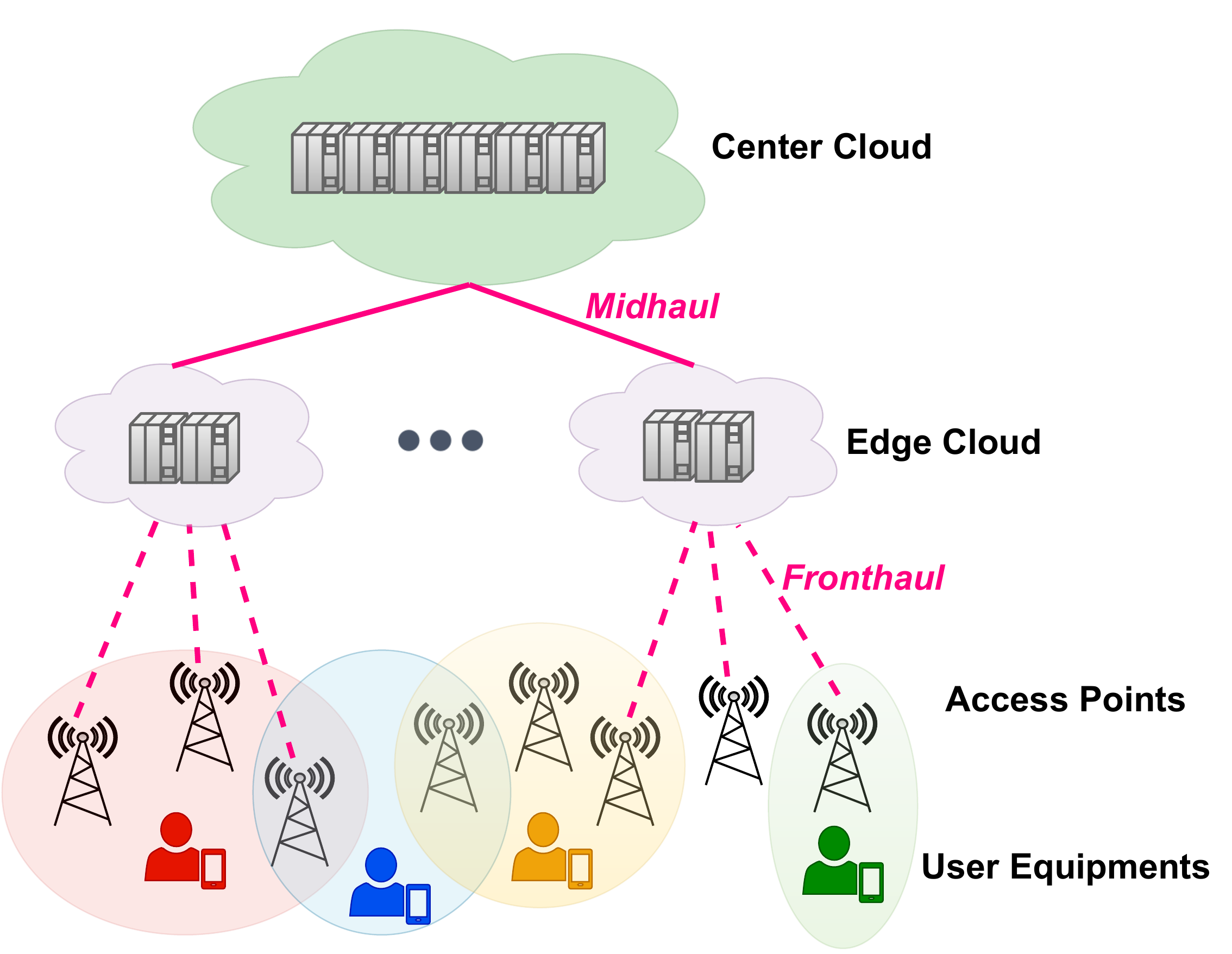}
\caption{The C-RAN architecture with cell-free massive MIMO functionality.}
\label{fig:cell-free}
\end{figure}

Cell-free massive MIMO is envisioned as a promising technology for beyond 5G systems due to the highly improved spectral and energy efficiency it would provide. As a natural consequence, not only the academia but also the industry has a great interest in cell-free massive MIMO, which is also named ``distributed MIMO'' or ``distributed massive MIMO'' by industrial researchers \cite{ref-cell-free}. It aims to guarantee almost uniformly great service to every user equipment in the coverage area by benefiting from joint transmission/reception and densely deployed low-cost access points with increased macro diversity. The physical-layer aspects such as receiver combining design, transmit precoding design, and power allocation algorithms in line with a futuristic scalable system design have now been well-established. For a scalable (in terms of signal processing complexity and fronthaul signaling load) cell-free massive MIMO system, an access point can only serve a finite number of user equipments. One service-oriented design option is the user-centric formation of the access points serving each user equipment according to their needs. As illustrated in Fig. ~\ref{fig:cell-free}, each user equipment is served by multiple access points with the preferable channel conditions, which are the ones in the colored shaded circular regions.   

The centralized computational processing unit and the fronthaul links between it and access points are two major layers in a practical cell-free massive MIMO operation envisioned to be built in 6G communication systems. When edge clouds are placed between the access points and the center cloud, as shown in Fig.~\ref{fig:cell-free}, the midhaul transport and the collaborative processing unit consisting of the edge and center cloud are the additional components in a cell-free network. Hence, the imperfections, limitations, and energy consumption should be analyzed from an end-to-end (from radio edge to the center cloud) perspective. Conducting an end-to-end study of a low-cost and energy-efficient cell-free massive MIMO implementation is critical to accelerating its practical deployment in 6G.

The network architecture of a cell-free massive MIMO system with access points connecting to central processing units via fronthaul links is entirely in line with the wave of cloudification in mobile communications networks. Hence, it is expected from the very beginning to envision prospective cell-free networks on top of a cloud radio access network (C-RAN). Virtualized C-RAN enables centralizing the digital units of the access points in an edge or central cloud with virtualization and computing resource-sharing capabilities. Going beyond virtualized C-RAN, the implementation options of cell-free massive MIMO have been discussed on top of open radio access networks (O-RAN) aiming for an intelligent, virtualized, and fully interoperable 6G architecture \cite{ref-O-RAN}. 

Fronthaul/midhaul transport technology is one of the vital components in the low-cost deployment of cell-free massive MIMO onto the legacy network. In a large-scale cell-free massive MIMO system, deploying a dedicated optical fiber link between each access point and the edge or central cloud would be highly costly and infeasible. The so-called ``radio stripes''-based fronthaul architecture developed by Ericsson reduces the cabling cost by sequentially integrating the access points into the shared fronthaul lines. When access points are distributed in a large area, other low-cost fronthaul transport technologies such as millimeter wave and terahertz wireless can both provide huge bandwidth and avoid costly wired fiber links. One other option is combined fiber-wireless fronthaul/midhaul transport to balance a trade-off between link quality and cost. In the latter method, the short-distance fronthaul links can be deployed wirelessly between each access point and its respective edge cloud. On the other hand, the midhaul transport from the edge to the center can benefit from extra-reliable fiber connections. Mitigating hardware impairments that naturally appear as a result of low-cost transceivers deployed at the access points and wireless fronthaul nodes is another critical aspect of the cell-free massive MIMO deployment on the legacy network.

In recent years, energy-saving techniques by mobile operators have gained more importance in reducing the environmental footprint and designing next-generation mobile communication systems in a green and sustainable way. Several works considered access point switching on/off methods in this research direction to save energy in a cell-free massive MIMO system. In addition, the virtualization and sharing of cloud and fronthaul/midhaul resources are crucial for minimizing total end-to-end energy consumption. At the end of the day, one should consider the limitations, energy consumption models, and the energy-saving mechanisms of digital units and processors in the edge and center cloud for the complete treatment of energy efficiency in a cell-free massive MIMO system.

\section{Artificial Intelligence for Massive MIMO}  

Massive MIMO technology powered by artificial intelligence (AI) can be applied to Industry 5.0. This technology can realize highly reliable real-time transmission of industrial 6G and other information with reliable human-computer interaction (HCI). Massive MIMO is a core technology of 5G. Still, the increase in the number of antennas has brought new challenges, significantly the rapid growth in the cost of channel estimation and feedback. Moreover, the accuracy of channel estimation and prediction needs to be improved. The application of AI technology is expected to solve the above problems. However, the above up-and-coming technologies have some issues. On the one hand, from the industry perspective, the main challenges are:

\begin{itemize}
    \item It is not easy to effectively control the difference between the training data set and the actual channel. The lack of generalization of AI algorithms may lead to a decline in system performance.
    \item Wireless AI data and applications have their unique characteristics. However, how to organically integrate the classic AI algorithms in image and voice processing with wireless data is still unclear.
    \item One of the characteristics of the Massive MIMO communication system applied to Industry 5.0 is that the communication scenarios are complex and changeable (indoor, outdoor, etc.), and the business forms are diverse. Therefore, making the wireless AI solution applicable to various communication scenarios and business forms under limited computing power is a significant challenge that the industry needs to overcome.
\end{itemize}

On the other hand, there are several interesting trends from the research perspective. First, applying machine learning into resource allocation has the potential to achieve low complexity implementation and decrease operational costs for massive MIMO. This strategy can improve spectral efficiency and energy efficiency, increase the number of users, and decrease energy consumption as well as the time delay. Second, using machine learning or deep learning for signal detection in massive MIMO has the potential to mitigate the high complexity issues seen in the conventional linear and non-linear detection methods. Third, AI can play a potential role in interference management for massive MIMO, such as determining and predicting the number of interference sources and strengths, and further mitigating the interference. Last but not least, with Massive MIMO expanding to more verticals, developing and deploying suitable segmented AI strategies for specific applications is critical.

\section{Massive MIMO-OFDM for High-Speed Applications}  

For massive MIMO, a very large number of antennas is used to either to reduce the multi-user interference (MUI), when spatially multiplexing several users, or to compensate the path loss when higher frequencies than microwave are used, such as the millimeter-waves (mm-Waves). Usually, a coherent demodulation scheme (CDS) is used in order to exploit MIMO-OFDM (orthogonal frequency-division multiplexing), where the channel estimation and the pre/post-equalization processes are complex and time-consuming operations, which require a considerable pilot overhead and increase the latency of the system. Moreover, new challenging scenarios are considered in 5G and beyond, such as high mobility scenarios (e.g., vehicular communications). The performance of the traditional CDS is even worse since reference signals cannot effectively track the fast variations of the channel with an affordable overhead. 

As an alternative solution, non-coherent demodulation schemes (NCDS) based on differential modulation combined with massive MIMO-OFDM have been proposed \cite{Chen-MIMO}. It is shown that even in the absence of reference signals, they can significantly outperform the CDS with a reduced complexity in high-speed scenarios, where no reference signals are required. In order to successfully implement the NCDS with the MIMO-OFDM system, some relevant details should be noted as follows. First, the high number of antennas is a key aspect to successfully deploy the NCDS. In the uplink, these antennas are used as spatial combiner capable of reducing the noise and self-interference produced by the differential modulation. In the downlink, beamforming is combined with NCDS in order to increase the coverage and spatially multiplex the different users. Then, the differential modulation should be mapped in the two-dimensional time-frequency resource grid of the OFDM symbol. Different schemes are proposed: time domain, frequency domain and hybrid domain, where the latter exhibits the best performance since it can minimize required signaling to a single pilot symbol for each transmitted burst. Last, on top of the MIMO-OFDM system, multiple users can be multiplexed in the constellation domain, which is an additional dimension to the existing spatial, time and frequency dimensions. At the transmitter, each user is choosing its own individual constellation, while at the receiver, the received joint constellation is a superposition of all individual constellations of each user. The overall performance in terms of bit error rate (BER) depends on the design of the received joint constellation, all chosen individual constellation and the mapped bits of each symbol. This non-convex optimization problem is solved using evolutionary computation, which is a subfield of artificial intelligence, capable of solving this kind of mathematical problems.

\begin{figure}[t!]
\centering
\includegraphics[trim={0mm 1mm 2mm 4mm},clip,width=0.42\textwidth]{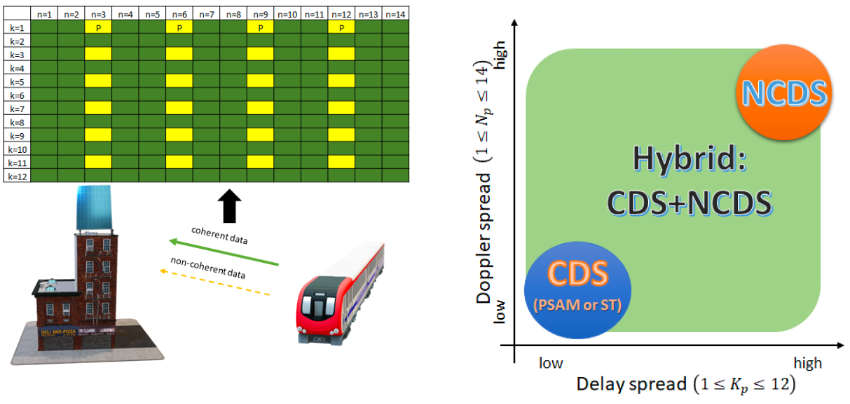}
\caption{Massive MIMO for high-speed applications, a design example of hybrid demodulation scheme.}
\label{fig:Vehicular}
\end{figure}

Finally, in those low or medium-mobility scenarios, a hybrid demodulation scheme (HDS) is proposed in \cite{Morales-MIMO}, which consists of replacing the traditional reference signals in CDS by a new differentially encoded data stream that can be non-coherently detected. The latter can be demodulated without the knowledge of the channel state information and subsequently used for the channel estimation. An design example is illustrated in Fig.~\ref{fig:Vehicular}. Consequently, HDS can exploit both the benefits of a CDS and NCDS to increase the spectral efficiency. It is outlined that the channel estimation is almost as good as CDS, while the BER performance and throughput are improved for different channel conditions with a very small complexity increase.

\section{Massive MIMO for Non-Terrestrial Communications}  
With 5G standardization phase 1 \& 2 finalized in 3GPP release 15 \& 16, the first half of 2022 has witnessed the third installment of the global 5G standard reaching the system design completion in 3GPP release 17 deemed as a continued expansion to 5G new devices and applications. In particular, 3GPP release 17 has introduced the 5G NR support for satellite communications which is one critical family member of the non-terrestrial networks (NTN). In fact, the concept of the NTN encompasses any network involving flying objects in either the air or space, and the NTN family therein includes satellite communication networks, high-altitude platform systems (HAPS) (including airplanes, balloons, and airships), and air-to-ground networks \cite{Lin-Space}.

As the focus of 3GPP NTN work, the satellite communication networks enable advanced features such as ubiquitous connectivity and coverage for remote/rural areas. Moreover, they include two distinct aspects, one focusing on satellite backhaul communications for application scenarios such as customer-premises equipment (CPE) and direct low data rate services for handhelds, while another one aims at adapting eMTC (enhanced Machine Type Communication) and NB-IoT (Narrowband Internet-of-Things) operation to satellite communications. Recent years have witnessed the unprecedented interest and prosperity in the Low Earth orbit (LEO) satellites enabled broadband access and services \cite{Huo-Space}. Among several major commercial players in the arena, namely Starlink (SpaceX), Kuiper (Amazon), OneWeb, Boeing, and Telestat, Starlink leads in terms of scale and dimension of satellite megaconstellation and number of service subscribers.   

There are several essential catalysts for accelerating the fast-booming spaceborne broadband access \cite{Huo-Space}, such as the launch cost decrease, private capitals, wide deployment of AI and cloud/edge computing, and high-performance satellite wireless and networking technologies. From wireless communications and a particular massive MIMO perspective, an overview of trends and challenges is presented as follows.

\begin{figure}[t!]
\centering
\includegraphics[trim={0mm 1mm 2mm 4mm},clip,width=0.46\textwidth]{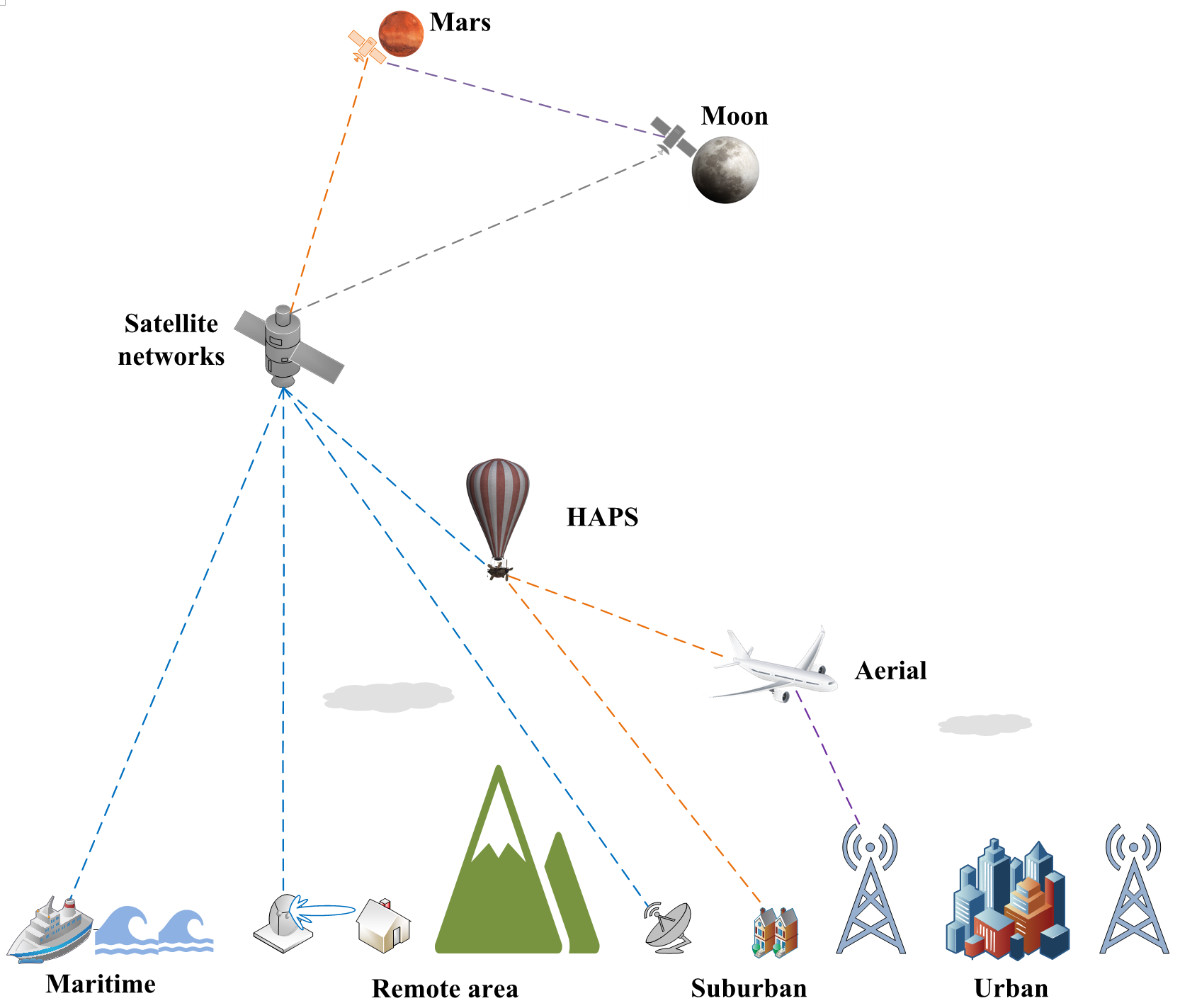}
\caption{Illustration of massive MIMO for non-terrestrial networks (NTNs).}\label{fig:NTN}
\end{figure}

\begin{itemize}
    \item Deploying and operating expanding satellite constellations below 2,000 km brings potential challenges of handling competition and facilitating coexistence. In order to minimize the propagation delay, many LEO megaconstellation builders may want to deploy satellites in orbits as low as possible, while the ITU adapts the ``First Come, First Served'' approach in the ITU cooperative system to access orbit/spectrum resources. With the space in the LEO becoming more crowded and collision incidents being reported \cite{Huo-Space}, regulations, strategies, and technologies are required to cope with increasing space traffic and handle safe deorbiting/disposal of satellites/spacecraft. With the escape velocity being at 7.8 km/s, tracking, localizing the satellites/spacecraft and further enabling collision-avoidance can be difficult even with contemporary AI-assisted sensing and detecting technologies.         
    
    \item Moreover, spectrum management is another critical aspect since the generally limited spectrum resource poses severe challenges. To facilitate 5G NR for NTN, 3GPP release 17 has investigated supporting satellites backhaul communication for CPEs and direct link to handhelds for low data rate services using sub-7 GHz S-band, while using frequency higher than 10 GHz will be studied in 3GPP release 18. In the meantime, the first-generation system of Starlink, or Gen1, has mainly used Ku-band and Ka-band for different types of links and transmission directions, and its second-generation (Gen2) will add V-band into it. Either sub-7 GHz S-band or Ku/Ka/V-band, to some extent, will overlap with some spectrum of ongoing 5G and future 6G systems, and/or other systems operating in these bands. Consequently, there can be interference and co-existence challenges among different systems and networks. In fact, both SpaceX and OneWeb have expressed concerns about the possible interference experienced by the non-geostationary orbit (NGSO) satellite internet if the terrestrial 5G uses 12 GHz band. Furthermore, supporting more satellite direct links to the user equipment (UE) using sub-7 GHz S-band also makes this interference challenge more pronounced. More studies of spectral resources (e.g. higher frequency bands) and spectrum management for spaceborne massive MIMO are expected.     
    
    \item Furthermore, there are various types of interferences that could emerge both within the same space network and among different space networks. For example, the in-band/out-band interference (or emission) can happen for user terminals (UTs) and ground stations of the same megaconstellation. When it comes to the situation of multiple space networks, satellite transmission of one megaconstellation could cause interference to the reception by UTs and ground stations belonging to other megaconstellations. Also, the UTs/ground stations transmission could interfere with the satellite(s) in different constellations. Conventionally, co-existing space networks need to share frequency allocations (both uplink and downlink) with each other to mitigate the interference, which is based on the coordination. However, more high-performance interference mitigation technologies are required for coping with more complicated situations in the future.         



    \item From a big-picture viewpoint, one the one hand, there is a trend that NGSO megaconstellations will support or efficiently co-work with GEO (geosynchronous Earth orbit) networks, HAPS, air-to-ground networks, drone networks, etc. Enabling such a greater NTN eco-system can bring more challenges of handling co-existence and competition. On the other hand, the space-enabled networks and massive MIMO will be expanded to and beyond the near-Earth space (NES) which is the space from the layers of the neutral terrestrial atmosphere (160-200 km) up to the lunar orbit (around 384,400 km). For example, NOKIA Bell Labs will deploy the first LTE network on the Moon for NASA's Artemis program. In the proposed Solar Communication and Defense Networks (SCADN) concept \cite{Huo-SCADN}, a massive MIMO sensing and communications framework based on an internet of a large number of spacecraft/satellites across the entire solar system enables early detection and mitigation of potential threats (e.g. asteroid/comet) to Earth and extra-terrestrial human bases, and also provides infrastructure facility to wireless connectivity within the solar system before/when human presence establishes on other celestial bodies. The very large propagation distances (between Earth and another celestial bodies) and delays pose severe challenges to the wireless sensing and communications, which requires more innovative solutions such as artificial intelligence, machine/deep learning, edge computing, edge AI, distributed and federated learning, etc.                      
\end{itemize}

To sum up, the massive MIMO technology has been fast extending the communicating and sensing capabilities of humanity beyond the terrain and even Earth, which will undoubtedly facilitate a more prosperous space era for all mankind.

\section{Conclusions}

This article presents a comprehensive overview of promising technology trends for massive MIMO on the evolving path to 6G. First, we conduct an overview of massive MIMO's recent standardization and research progress. Then we focus on IRS/IOS technologies that can enable/cost-efficient massive MIMO communications. Furthermore, we envision the challenges of using IRS/IOS for facilitating and enhancing the localization and sensing capabilities in massive MIMO. Next, we investigate the ultra-massive MIMO at THz frequencies and unveil several impairments that affect the system design. Then we present and analyze the cell-free massive MIMO architecture which can boost the spectral and energy efficiency of wireless systems and networks. In addition, the challenges and trends of AI for massive MIMO are discussed in depth. Meanwhile, future massive MIMO will enable and strengthen more critical vertical applications. Therefore, the massive MIMO-OFDM-enabled high-speed communications is surveyed and presented with some designed examples. Finally, we carefully present and analyze the current and future trends of massive MIMO communications for non-terrestrial networks, particularly near-Earth space and inter-planetary applications.

\section*{Acknowledgments}
We express sincere thanks to the IEEE Future Networks Massive MIMO Working Group, and the Organizing Committee of the IEEE Future Networks Second Massive MIMO Workshop. All co-authors contributed equally in this manuscript.

\newpage


\vfill

\end{document}